\documentclass[12pt,a4]{article}
\usepackage{amsmath}
\usepackage{bm}
\usepackage{amsfonts}
\usepackage{amssymb}
\usepackage{graphics}
\usepackage[normal]{caption2}
\usepackage{subfigure}
\usepackage{rotating}
\usepackage{citesort}
\def\be{\begin{equation}}
\def\ee{\end{equation}}
\def\ba{\begin{array}}
\def\ea{\end{array}}
\def\bea{\begin{eqnarray}}
\def\eea{\end{eqnarray}}

\begin{document}
\baselineskip 20pt \setlength\tabcolsep{2.5mm}
\renewcommand\arraystretch{1.5}
\setlength{\abovecaptionskip}{0.1cm}
\setlength{\belowcaptionskip}{0.5cm}
\begin{center} {\large\bf Sensitivity of the transverse flow towards symmetry energy}\\
\vspace*{0.4cm}
{\bf Sakshi Gautam$^a$}, {\bf Aman D. Sood$^b$}, {\bf Rajeev K. Puri$^a$}\footnote{Email:~rkpuri@pu.ac.in}{\bf and J. Aichelin $^b$}\\
$^a${\it  Department of Physics, Panjab University, Chandigarh
-160 014, India.\\} $^b${\it  SUBATECH, Laboratoire de Physique
Subatomique et des Technologies Associ\'{e}es, Universit\'{e} de
Nantes - IN2P3/CNRS - EMN 4 rue Alfred Kastler, F-44072 Nantes,
France.\\}
\end{center}
We study the sensitivity of transverse flow towards
 symmetry energy in the Fermi energy region as well as at high energies.
 We find that transverse flow is sensitive to symmetry energy as
 well as its density dependence in the Fermi energy region. We also show that the transverse flow can address
 the symmetry energy at densities about twice the saturation density, however it shows the insensitivity towards
 the symmetry energy at densities $\rho/\rho_{0}$ $>$ 2. The
 mechanism for the sensitivity of transverse flow towards symmetry
 energy as well as its density dependence is also discussed.


\newpage
\baselineskip 20pt
\section{Introduction}
The heavy-ion collisions (HIC) are the only method presently
available in the laboratory to produce large volume of excited
nuclear matter. The production of such state is essential to
investigate not only the gross characteristics of nuclear matter
but also in exploring the explosion mechanism of supernova and
cooling rate of neutron stars. After about three decades of
extensive efforts in both nuclear experiments and theoretical
calculations, equation of state (EOS) of isospin symmetric matter
is well understood by experiments of collective flow \cite{sch}
and subthreshold kaon production \cite{klak93,kaon1}. Nowadays,
the nuclear EOS of asymmetric nuclear matter has attracted a lot
of attention. The EOS of asymmetric nuclear matter can be
described approximately by
\begin{equation}
E(\rho, \delta) = E_{0}(\rho,
\delta=0)+E_{\textrm{sym}}(\rho)\delta^{2}
\end{equation}
where
 $\delta$ = $\frac{\rho_{n}-\rho_{p}}{\rho_{n}+\rho_{p}}$ is isospin asymmetry,
  E$_{0}$($\rho$, $\delta$) is the energy of pure symmetric nuclear matter and E$_{\textrm{sym}}$($\rho$)
  is the symmetry energy with E$_{\textrm{sym}}$($\rho$$_{0}$) = 32 MeV is the symmetry energy at normal
  nuclear matter density. The symmetry energy is E($\rho$,1) - E$_{0}$($\rho$,0), ie. the difference
  of the energy per nucleon between pure neutron matter and symmetric nuclear matter.
  The symmetry energy is important not only to the nuclear physics community
  as it sheds light on the structure of radioactive nuclei, reaction dynamics induced
by rare isotopes but also to astrophysicists since it acts as a
probe for understanding the evolution of massive stars and the
supernova explosion \cite{kubis07}. The existing and upcoming
radioactive ion beam (RIB) facilities led a way in understanding
nuclear symmetry energy. Experimentally, symmetry energy is not a
directly measurable quantity and has to be extracted from
observables which are related to symmetry energy. Therefore, a
crucial task is to find such observables which can shed light on
symmetry energy. A large number of studies on the symmetry energy
of nuclear matter have been done during the past decade
\cite{zhang05,qli06,tsang09,bali97,baran98,li02,li04}. These
studies reveal that in heavy-ion collisions induced by
neutron-rich nuclei, the effect of nuclear symmetry energy can be
studied via the pre equilibrium \emph{n/p} ratio
\cite{zhang05,qli06,tsang09}, isospin fractionation
\cite{bali97,baran98}, \emph{n-p }differential transverse flow
\cite{li02,li04} and so on.  These observables have their relative
importance depending upon the region of density one wants to
explore. For e.g. below saturation density (0.3 $\rho_{0} \leq
\rho \leq \rho_{0}$), observables like fragment yield, isoscaling
parameter, isospin diffusion, double \textit{n}-\emph{p} ratio
have been used to extract symmetry energy. On the other extreme,
$\pi^{+}$/$\pi^{-}$ ratio, relative and differential collective
flow between triton/He$^{3}$, \emph{n}-\emph{p} differential
collective flow act as probe of symmetry energy at high densities.
In the low density region, Shetty \emph{et al.} \cite{shetty04}
extracted the symmetry energy by comparing the isoscaling
parameter from $^{40}$\textrm{Ar},
$^{40}$\textrm{Ca}+$^{58}$\textrm{Fe}, $^{58}$\textrm{Ni} and
$^{58}$\textrm{Fe}, $^{58}$\textrm{Ni}+$^{58}$\textrm{Fe},
$^{58}$\textrm{Ni} reactions with dynamical Antisymmetrized
Molecular Dynamics (AMD) calculations \cite{ono03} and found it to
be of the form E$_{\textrm{sym}} = 31.6
(\frac{\rho}{\rho_{0}})^{\gamma}$, with $\gamma$ = 0.69. Famiano
 \emph{et al}. \cite{fami06} studied the symmetry energy by comparing the experimental
double neutron to proton ratio in
$^{112}$\textrm{Sn}+$^{112}$\textrm{Sn} and
$^{124}$\textrm{Sn}+$^{124}$\textrm{Sn} reactions
Isospin-dependent Boltzmann-Uehling-Uhlenbeck (IBUU) calculations
\cite {li97} and obtained the form E$_{sym} = 32
(\frac{\rho}{\rho_{0}})^{\gamma}$, $\gamma$ = 0.5. Recently Tsang
\emph{et al}. \cite{tsang09} compared the isospin diffusion and
n-p double ratio for $^{124}$\textrm{Sn}+$^{112}$\textrm{Sn}
reaction with Isospin-dependent Quantum Molecular Dynamics (IQMD)
calculations \cite{zhang08} and obtained a similar form of
symmetry energy with $\gamma$ = 0.4-1.05. The situation is worse
at higher densities. The results are model dependent and
contradicting also. The FOPI collaboration at GSI studied
$\pi^{+}$/$\pi^{-}$ ratio in
$^{40}$\textrm{Ca}+$^{40}$\textrm{Ca},
$^{96}$\textrm{Ru}+$^{96}$\textrm{Ru},
$^{96}$\textrm{Zr}+$^{96}$\textrm{Zr}, and
$^{197}$\textrm{Au}+$^{197}$\textrm{Au} reactions \cite{reis07}. A
comparison of this data \cite{xiao09} with IBUU calculations
showed a softer form of the density dependence of symmetry energy,
which is in contrast to those obtained from the low density
studies where stiffer form of symmetry energy reproduced the data
well. $\pi^{+}$/$\pi^{-}$ ratio was also compared with Improved
Quantum Molecular Dynamics (ImQMD) calculations by Feng \emph{et
al}. \cite{feng10} and favored a stiffer form of symmetry energy
and in contradiction with IBUU calculations. At densities higher
than saturation density, collective flow has also been proposed as
a novel mean to probe the high density behavior of symmetry energy
\cite{li02}. In this paper, we aim to see the sensitivity of
collective transverse in-plane flow towards symmetry energy and
also to see the effect of different density dependence of symmetry
energy on the same. The various forms of symmetry energy used in
present study are: E$_{\textrm{sym}} \propto F_{1}(u)$,
E$_{\textrm{sym}} \propto F_{2}(u)$, and E$_{\textrm{sym}} \propto
F_{3}(u)$, where \emph{u} = $\frac{\rho}{\rho_{0}}$, F$_{1}(u)
\propto u$, F$_{2}(u) \propto u^{0.4}$, F$_{3}(u) \propto u^{2}$,
and F$_{4}$ represents calculations without symmetry energy. The
different density dependences of symmetry energy are shown in fig.
1. The various lines are explained in the caption of the fig. 1.
Section 2 describes the model in brief. Section 3 explains the
results and discussion and Sec. 4 summarizes the results.

\begin{figure}[!t] \centering
 \vskip -1cm
\includegraphics[width=12cm]{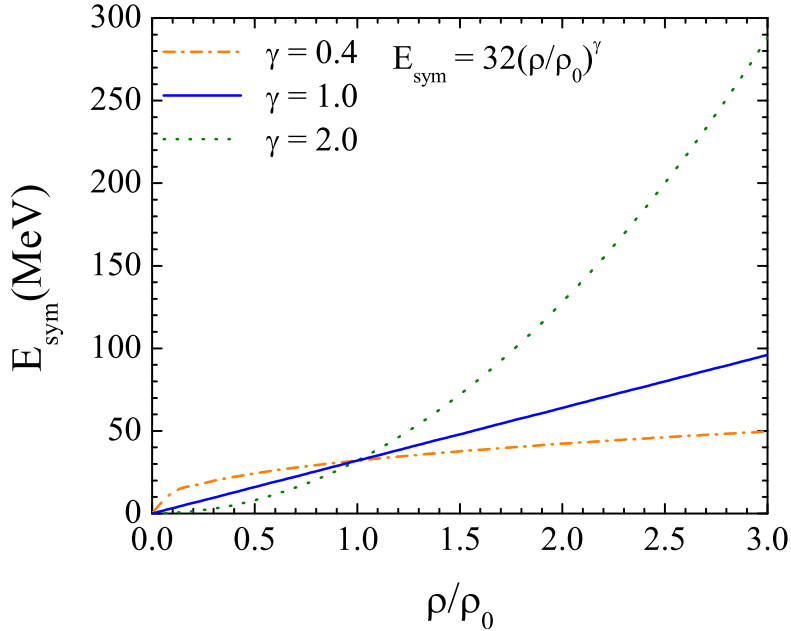}
\caption{(Color online) The density dependence of symmetry energy
for various forms of symmetry energy : F$_{1}$ ($\gamma$ = 1.0)
(solid), F$_{2}$ ($\gamma$ = 0.4) (dashed), and F$_{3}$ ($\gamma$
= 2.0) (dotted).}\label{fig2}
\end{figure}

 \par
 \section{The model}
 The present study is carried out within the framework of
isospin-dependent quantum molecular dynamics (IQMD) model
\cite{hart98}. The IQMD model treats different charge states of
nucleons, deltas, and pions explicitly, as inherited from the
Vlasov-Uehling-Uhlenbeck (VUU) model. The IQMD model has been used
successfully for the analysis of a large number of observables
from low to relativistic energies. The isospin degree of freedom
enters into the calculations via symmetry potential, cross
sections, and Coulomb interaction.
 \par
 In this model, baryons are represented by Gaussian-shaped density distributions
\begin{equation}
f_{i}(\vec{r},\vec{p},t) =
\frac{1}{\pi^{2}\hbar^{2}}\exp(-[\vec{r}-\vec{r_{i}}(t)]^{2}\frac{1}{2L})
\times \exp(-[\vec{p}- \vec{p_{i}}(t)]^{2}\frac{2L}{\hbar^{2}})
 \end{equation}
 Nucleons are initialized in a sphere with radius R = 1.12 A$^{1/3}$ fm, in accordance with liquid-drop model.
 Each nucleon occupies a volume of \emph{h$^{3}$}, so that phase space is uniformly filled.
 The initial momenta are randomly chosen between 0 and Fermi momentum ($\vec{p}$$_{F}$).
 The nucleons of the target and projectile interact by two- and three-body\textrm{ Skyrme} forces, \textrm{Yukawa} potential,and \textrm{Coulomb} interactions. In addition to the use of explicit charge states of all baryons
and mesons, a symmetry potential between protons and neutrons
 corresponding to the Bethe-Weizs\"acker mass formula has been included. The hadrons propagate using Hamilton equations of motion:
\begin {eqnarray}
\frac{d\vec{{r_{i}}}}{dt} = \frac{d\langle H
\rangle}{d\vec{p_{i}}};& & \frac{d\vec{p_{i}}}{dt} = -
\frac{d\langle H \rangle}{d\vec{r_{i}}}
\end {eqnarray}
 with
\begin {eqnarray}
\langle H\rangle& =&\langle T\rangle+\langle V \rangle
\nonumber\\
& =& \sum_{i}\frac{p^{2}_{i}}{2m_{i}} + \sum_{i}\sum_{j>i}\int
f_{i}(\vec{r},\vec{p},t)V^{\textrm{ij}}(\vec{r}~',\vec{r})
 \nonumber\\
& & \times f_{j}(\vec{r}~',\vec{p}~',t) d\vec{r}~ d\vec{r}~'~
d\vec{p}~ d\vec{p}~'.
\end {eqnarray}
 The baryon potential V$^{\textrm{ij}}$, in the above relation, reads as
 \begin {eqnarray}
  \nonumber V^{\textrm{ij}}(\vec{r}~'-\vec{r})& =&V^{\textrm{ij}}_{\textrm{Skyrme}} + V^{\textrm{ij}}_{\textrm{Yukawa}} +
  V^{\textrm{ij}}_{\textrm{Coul}} + V^{\textrm{ij}}_{\textrm{sym}}
    \nonumber\\
   & =& [t_{1}\delta(\vec{r}~'-\vec{r})+t_{2}\delta(\vec{r}~'-\vec{r})\rho^{\gamma-1}(\frac{\vec{r}~'+\vec{r}}{2})]
   \nonumber\\
   &  & +t_{3}\frac{\exp(|(\vec{r}~'-\vec{r})|/\mu)}{(|(\vec{r}~'-\vec{r})|/\mu)}+
    \frac{Z_{i}Z_{j}e^{2}}{|(\vec{r}~'-\vec{r})|}
   \nonumber \\
      &  & +t_{4}\frac{1}{\varrho_{0}}T_{\textrm{3i}}T_{\textrm{3j}}\delta(\vec{r_{i}}~'-\vec{r_{j}}).
 \end {eqnarray}
Here \emph{Z$_{i}$} and \emph{Z$_{j}$} denote the charges of
\emph{ith} and \emph{jth} baryon, and \emph{T$_{3i}$} and
\emph{T$_{3j}$} are their respective \emph{T$_{3}$} components
(i.e., $1/2$ for protons and $-1/2$ for neutrons). The
parameters\emph{ $\mu$} and \emph{t$_{1}$,...,t$_{4}$} are
adjusted to the real part of the nucleonic optical potential.
 For the density dependence of  the nucleon optical potential, standard \textrm{Skyrme} type parametrization is employed.

\begin{figure}[!t] \centering
 \vskip 1cm
\includegraphics[angle=0,width=8cm]{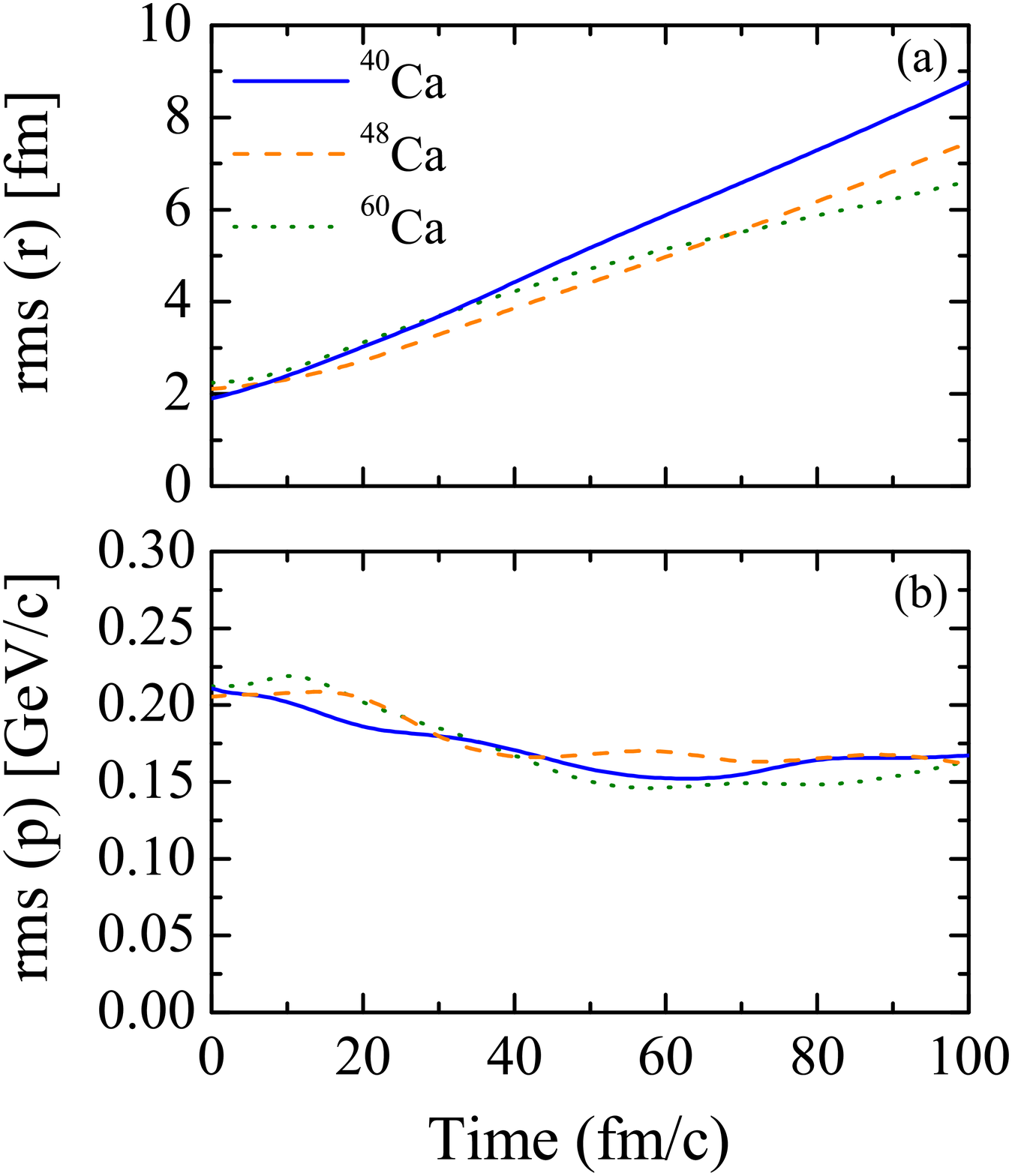}
 \vskip -0cm \caption{ (Color online) Time evolution of root mean square
 radius of a single $^{40}$\textrm{Ca}, $^{48}$\textrm{Ca} and $^{60}$\textrm{Ca} nuclei in coordinate (top panel) and momentum space (bottom) obtained
 with IQMD for EOS used in the present study for
 symmetry energy F$_{1}$(u).} \label{fig1}
\end{figure}

\begin{figure}[!t] \centering
\vskip 0.5cm
\includegraphics[angle=0,width=12cm]{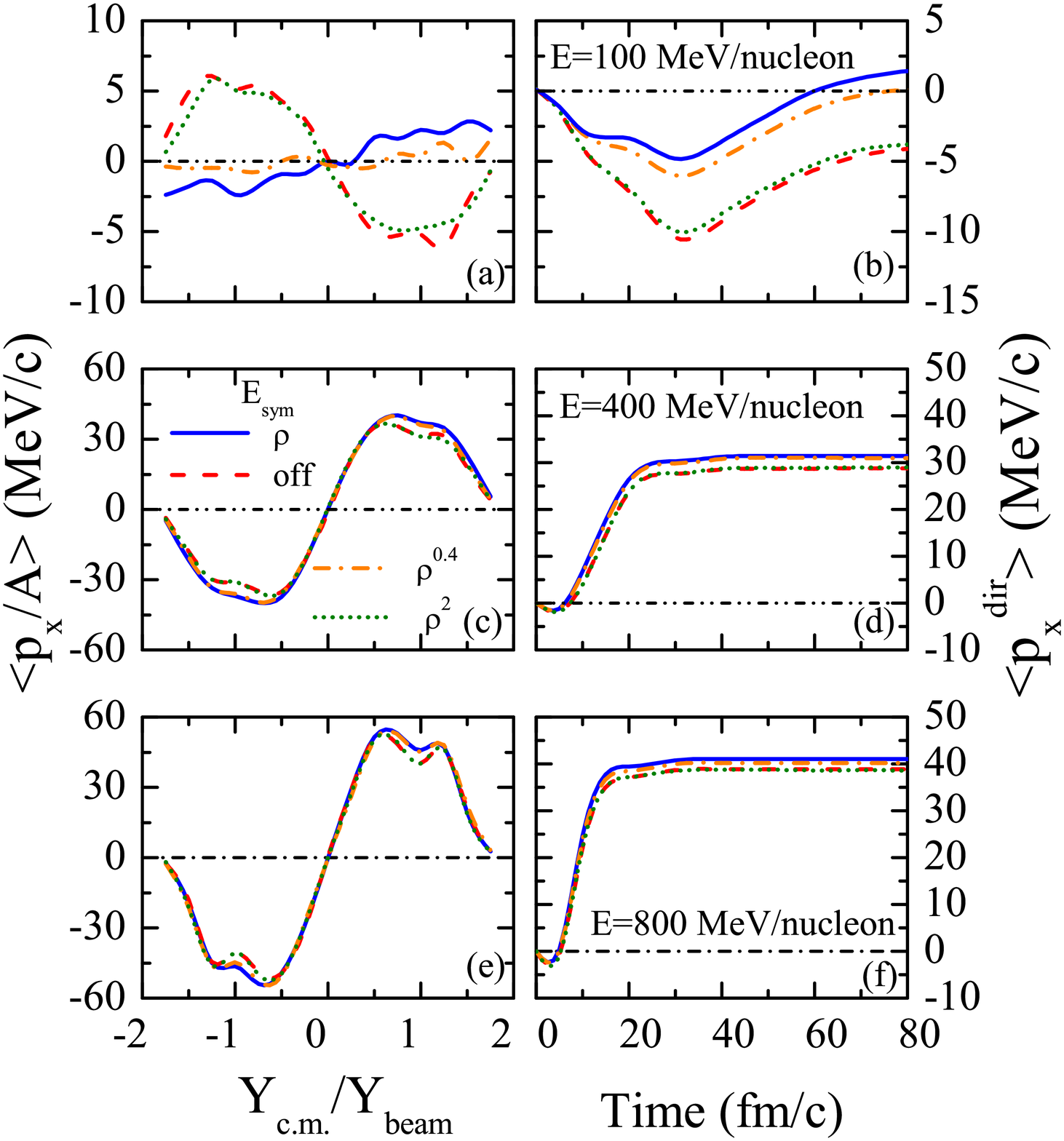}
\vskip 0.5cm \caption{(Color online) Left panel:
$<\frac{p_{x}}{A}>$ as a function of
Y$_{\textrm{c.m.}}/Y_{\textrm{beam}}$ for different energies of
100, 400, and 800 MeV/nucleon for different forms of symmetry
energy. Right panel: The time evolution of
$<p_{x}^{\textrm{dir}}>$ at 100, 400, and 800 MeV/nucleon for
different forms of symmetry energy at
b/b$_{\textrm{max}}$=0.2-0.4. Lines are explained in the
text.}\label{fig4}
\end{figure}

\begin{figure}[!t] \centering
 \vskip -1cm
\includegraphics[width=12cm]{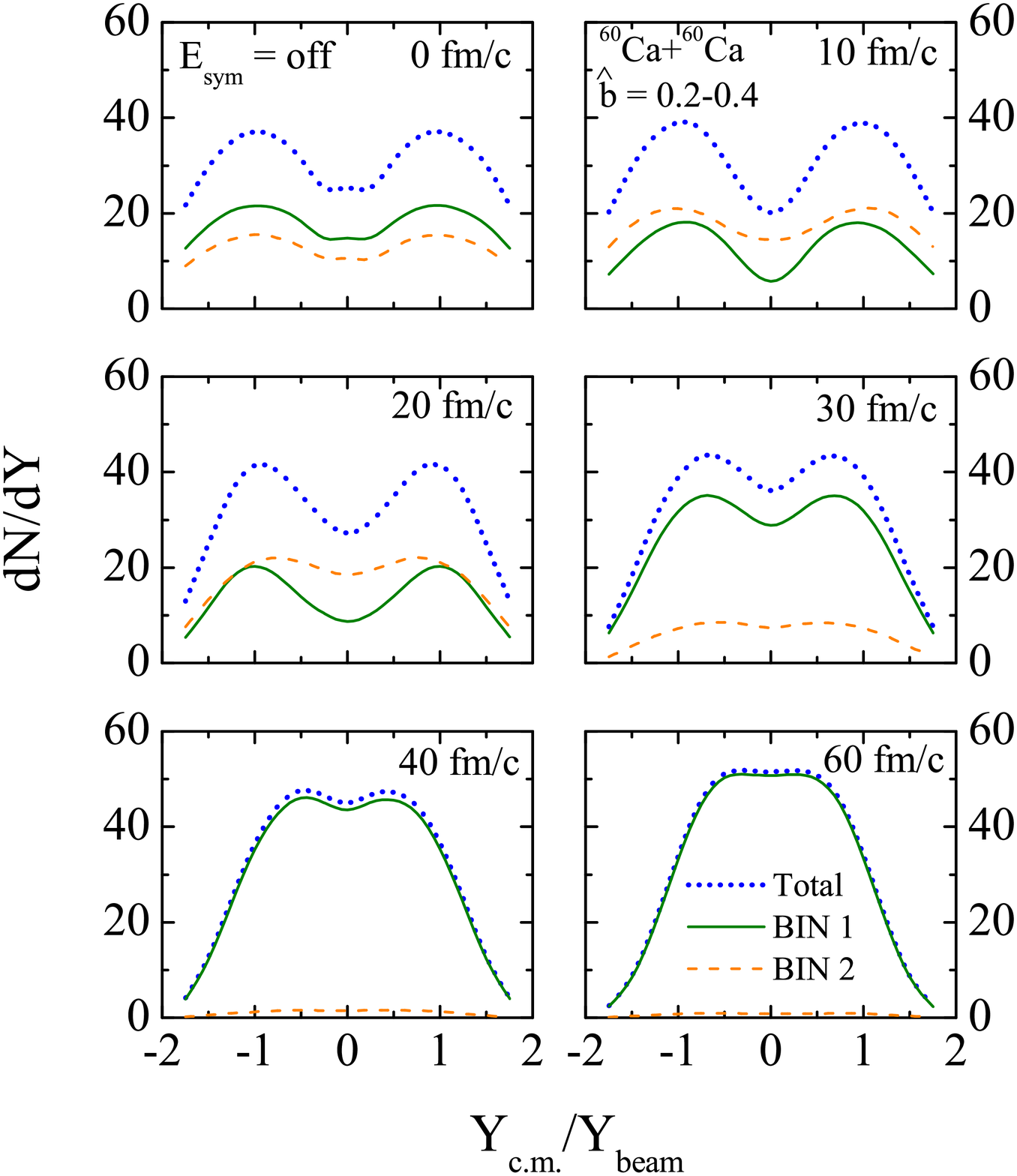}
\caption{(Color online) The time evolution of rapidity
distribution for the calculations with no symmetry energy for
various bins at b/b$_{\textrm{max}}$=0.2-0.4. Lines are explained
in the text.}\label{fig2}
\end{figure}

\par
 \section{Results and discussion}
We simulate several thousands events for the neutron-rich system
of  $^{48}\textrm{Ca}$+$^{48}\textrm{Ca}$  and
$^{60}\textrm{Ca}$+$^{60}\textrm{Ca}$ at energies of 100, 400, and
800 MeV/nucleon at impact parameter of
b/b$_{\textrm{max}}$=0.2-0.4. We use a soft equation of state and
isospin- and energy-dependent cross section reduced by
  20$\%$, i.e. $\sigma$ = 0.8 $\sigma_{nn}^{\textrm{free}}$.
The details about the elastic and inelastic cross sections for
proton-proton and proton-neutron collisions can be found in
\cite{hart98,cug}. The cross section for neutron-neutron
collisions is assumed to be equal to the proton-proton cross
section.
\par
Since $^{60}$\textrm{Ca} has a very high asymmetry, so to ensure
the stability of the nuclei in the present study, we display in
fig. 2 the time evolution of root mean square radius of single
nucleus of $^{40}$\textrm{Ca} (solid line), $^{48}$\textrm{Ca}
(dashed), and $^{60}$\textrm{Ca} (dotted) in the coordinate (fig.
2(a)) and momentum space (fig. 2(b)). The results are displayed
for nuclei intialized with symmetry energy F$_{1}$(u). We find
that the stability is of same order for all the three nuclei.

There are several methods used in the literature to define the
nuclear transverse in-plane flow. In most of the studies, one uses
($p_{x}/A$) plots where one plots ($p_{x}/A$) as a function of
$Y_{\textrm{c.m.}}/Y_{\textrm{beam}}$. Using a linear fit to the
slope, one can define the so-called reduced flow (F).
Alternatively, one can also use a more integrated quantity
``directed transverse in-plane flow
$\langle{p_{x}^{\textrm{dir}}}\rangle$'' which is defined as \cite
{hart98}:
\begin{equation}
\langle p_{x}^{\textrm{dir}}\rangle~=~\frac{1}{A}\sum_i {\rm
sign}\{Y(i)\}~p_{x}(i),
\end{equation}
where $Y(i)$ and $p_{x}(i)$ are the rapidity distribution and
transverse momentum of the $ith$ particle. In this definition, all
rapidity bins are taken into account. It, therefore, presents an
easier way of measuring the in-plane flow rather than complicated
functions such as ($p_{x}/A$) plots.
\par
 In fig. 3 we display the
$<\frac{p_{x}}{A}>$ as a function of
Y$_{\textrm{c.m.}}/Y_{\textrm{beam}}$ at final time (left panels)
and time evolution of $<p_{x}^{\textrm{dir}}>$ (right panels)
calculated at 100 (top panel), 400 (middle) and 800 MeV/nucleon
(bottom) for different density dependence of symmetry energy.
Solid, dash dotted, and dotted lines represent the symmetry energy
proportional to $\rho$, $\rho^{0.4}$ and $\rho^{2}$ whereas dashed
line represents calculations without symmetry energy. Comparing
left and right panels in fig. 3, we find that both the methods
show similar behavior to symmetry energy. For eg. at incident
energy of 100 MeV/nucleon for E$_{\textrm{sym}} \propto
\rho^{0.4}$, $<p_{x}^{\textrm{dir}}>$ = 0. Similarly, the slope of
$<\frac{p_{x}}{A}>$ at midrapidity is zero. We also find that the
transverse momentum is sensitive to symmetry energy as well as its
density dependence F$_{1}(u)$, F$_{2}(u)$ and F$_{3}(u)$ in the
low energy region (100 MeV/nucleon). At energies above Fermi
energy, both the methods show insensitivity to the different
symmetry energies. This is because the repulsive nn scattering
dominates the mean field at high energies.
\par
 To
understand the sensitivity of transverse momentum to the symmetry
energy as well as its density dependence in the Fermi energy
region, we calculate the transverse flow as well as rapidity
distribution of particles having $\frac{\rho}{\rho_{0}}$ $<$ 1
(denoted as bin 1) and particles having $\frac{\rho}{\rho_{0}}$
$\geq$ 1 (bin 2) separately at all the time steps. Since both the
methods show similar behavior to symmetry energy, so for
simplicity the following discussions will be in terms of
$<p_{x}^{\textrm{dir}}>$.

\par
In fig. 4 we display the rapidity distributions at 100 MeV/nucleon
of all the particles (dotted line), particles corresponding to bin
1 (solid) and to bin 2 (dashed) at 0, 10, 20, 30, 40, and 60 fm/c.
We have calculated rapidity distributions for different forms of
symmetry energy used in this paper. We find that it is insensitive
towards the symmetry energy \cite{indra,sanjeev}. In fig. 4, we
display the rapidity distribution calculated without symmetry
energy. During the initial stages we see the two Gaussians at
projectile and target rapidities for all the 3 bins. The peaks of
Gaussians will be more prominent at higher energies. The interest
for our discussion is in bin 1 and bin 2. During the start of the
reaction (0 fm/c) more number of particles lie in bin 1 i.e. more
number of particles have $\frac{\rho}{\rho_{\textrm{0}}}$ $<$ 1.
As the nuclei begin to overlap, the density increases in the
overlap zone. Now, the number of particles increases in bin 2 (at
10 fm/c). From 10 fm/c to 20 fm/c, the number of particles keep on
increasing in bin 2 at midrapidity, i.e particles from large
rapidity keep on shifting to the bin 2 in the midrapidity region.
This is expected since at incident energies in the Fermi energy
region dynamics is governed by the attractive mean field. The
dominance of attractive mean field will prompt the deflection of
particles into negative angles i.e. towards participant zone.
After 20-30 fm/c, the expansion phase of the reaction begins and
number of particles keep on increasing in bin 1 and by 60 fm/c
most of the particles lie in bin 1.

\begin{figure}[!t] \centering
 \vskip -1cm
\includegraphics[width=12cm]{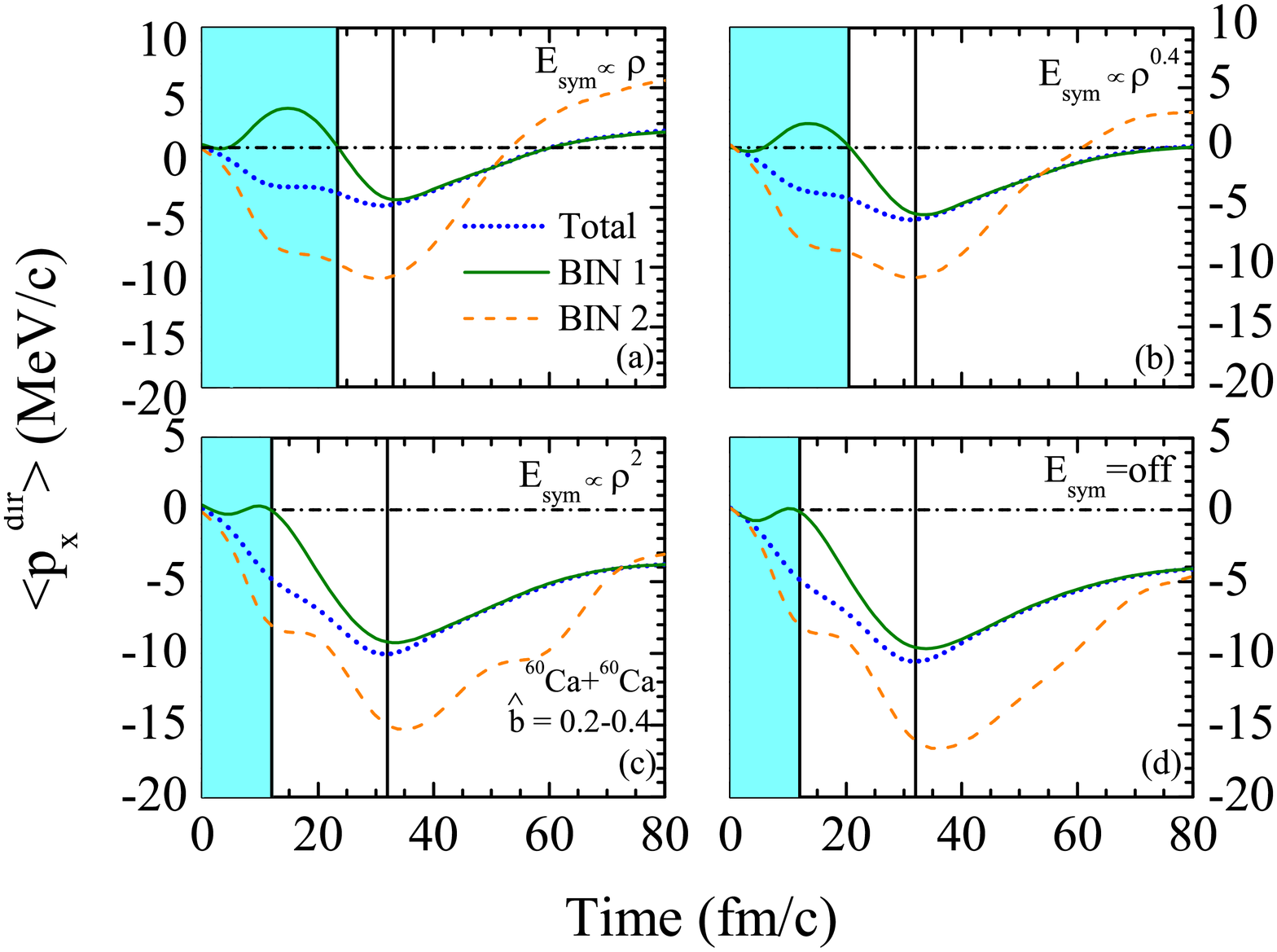}
\caption{(Color online) The time evolution of
$<p_{x}^{\textrm{dir}}>$ for different forms of symmetry energy
for different bins at b/b$_{\textrm{max}}$=0.2-0.4 . Lines have
same meaning as in fig. 4.}\label{fig2}
\end{figure}

 \par
In fig. 5 we display the time evolution of
$<p_{x}^{\textrm{dir}}>$ for different symmetry energies used in
this paper at 100 MeV/nucleon for particles lying in the different
bins. Lines have the same meaning as in fig. 4. Panels a, b, and c
are for E$_{\textrm{sym}} \propto \rho, \rho^{0.4}$, and
$\rho^{2}$, respectively. Panel d is for calculations without
symmetry energy. The total $<p_{x}^{\textrm{dir}}>$ is negative
during the initial stages and keep on decreasing till 30 fm/c
which indicates dominance of attractive interaction. In panels a
and b, it becomes positive whereas in panel c and d it remains
negative during the course of the reaction. If we look at
$<p_{x}^{\textrm{dir}}>$ of particles lying in bin 1 for F$_{1}
(u)$ (fig. 5a) and F$_{2} (u)$ (fig. 5b) in the time interval 0 to
about 20-25 fm/c, we see that it remains positive. It increases
with time upto 15 fm/c and reaches a peak value. This is because
in the spectator region (where high rapidity particles lies) the
repulsive symmetry energy will accelerate the particles away from
the overlap zone in the transverse direction. After 15 fm/c,
$<p_{x}^{\textrm{dir}}>$ (of particles in bin 1) begins to
decrease. This is because these particles will now be attracted
towards the central dense zone. As shown in fig. 4, from 10 to 20
fm/c particles in bin 2 keep on increasing in the midrapidity
region. In case of F$_{1} (u)$ and F$_{2} (u)$, particles which
enter the central dense zone (bin 2) have already a high positive
value of $<p_{x}^{\textrm{dir}}>$ (i.e. going away from dense
zone). So attractive mean field have to decelerate the particles
first, make them stop and then accelerate the particles back
towards the overlap zone. At about 20-25 fm/c particles from bin 1
have zero $<p_{x}^{\textrm{dir}}>$ (see shaded area in fig. 5a and
5b). Up to 30 fm/c, particles feel the attractive mean field
potential after which the high density phase is over, i.e. in case
of F$_{1} (u)$ and F$_{2} (u)$ between 0-30 fm/c  particles from
bin 1 are accelerated towards the overlap zone only for a short
time interval of about 5 fm/c, whereas for the case of F$_{3} (u)$
(fig. 5c) and F$_{4}$ (fig. 5d) between 0-30 fm/c, particles from
bin 1 are accelerated towards the overlap zone for a longer time
interval of about 20 fm/c between 10-30 fm/c. Moreover, the
$<p_{x}^{\textrm{dir}}>$ of particles lying in the bin 1 (for
F$_{3} (u)$ and F$_{4}$) follows a similar trend. This is because,
for $\rho/\rho_{0} < $ 1 the strength of symmetry energy F$_{3}
(u)$ will be small and so there will be less effect of symmetry
energy on the particles which is evident from fig. 5c where one
sees that the $<p_{x}^{\textrm{dir}}>$ remains about zero during
the initial stages between zero to about 10 fm/c.

\begin{figure}[!t] \centering
\vskip 0.5cm
\includegraphics[angle=0,width=10cm]{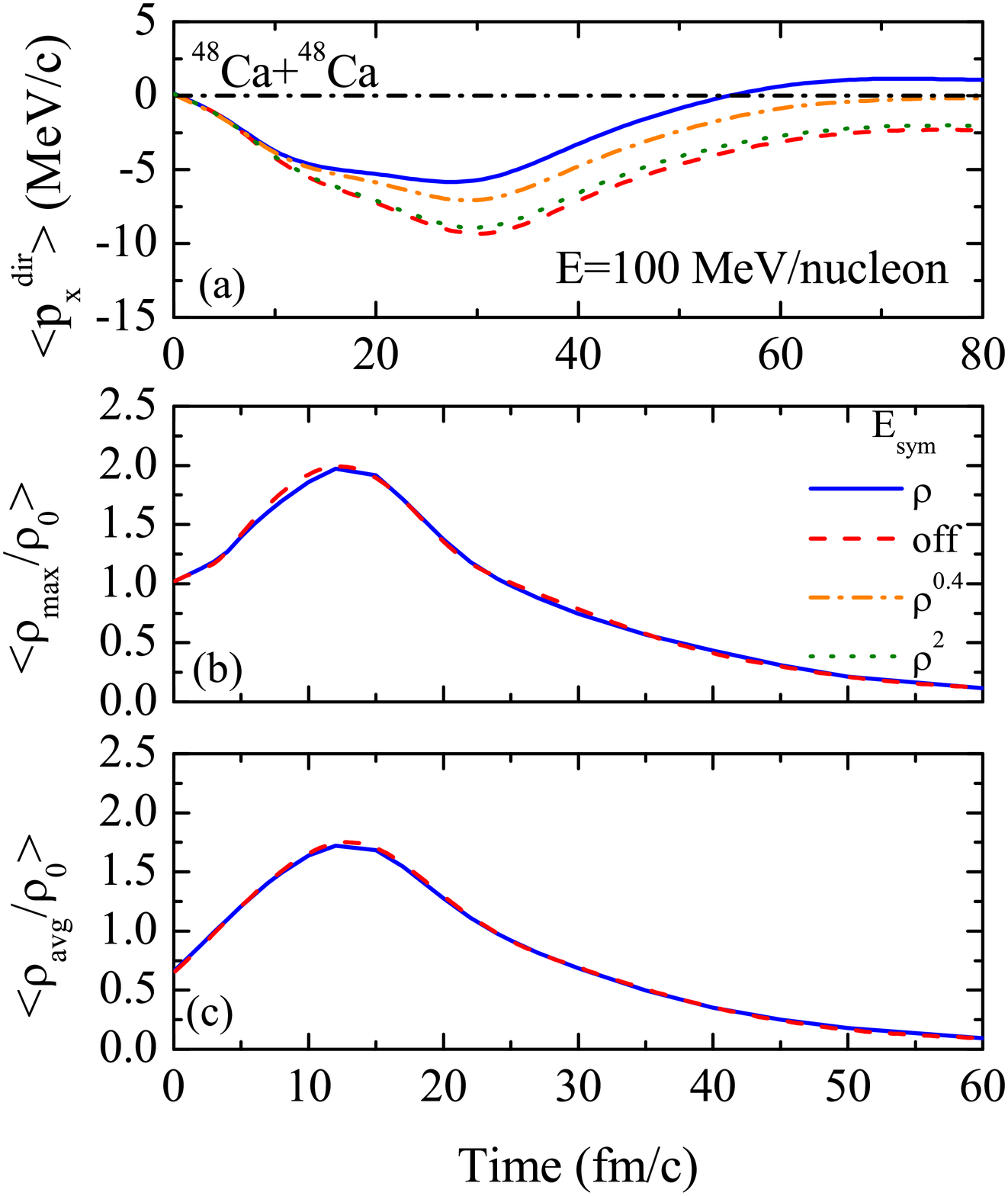}
\vskip 0.5cm \caption{(Color online) (a) The time evolution of
$<p_{x}^{\textrm{dir}}>$ at 100 MeV/nucleon for different forms of
symmetry energy for $^{48}$\textrm{Ca}+$^{48}$\textrm{Ca}. (b) The
time evolution of $<\rho_{\textrm{max}}/\rho_{\textrm{0}}>$ and
(c) $<\rho_{\textrm{avg}}/\rho_{\textrm{0}}>$ for F$_{1} (u)$ and
F$_{4}$. Lines have same meaning as in fig. 3.}\label{fig4}
\end{figure}

\par
 The $<p_{x}^{\textrm{dir}}>$ due to particles in bin 2 (dashed line) decreases in a very similar
 manner for all the four different symmetry energies between 0-10 fm/c. Between 10-25 fm/c, $<p_{x}^{\textrm{dir}}>$ for
 F$_{3} (u)$
 and F$_{4}$ decreases more sharply as compared to in
case of F$_{1} (u)$ and F$_{2} (u)$. This is because in this time
interval particles from bin 1 enters into bin 2. As discussed
earlier, $<p_{x}^{\textrm{dir}}>$ of particles entering bin 2 from
bin 1 in case of F$_{1} (u)$ and F$_{2} (u)$ will be less negative
due to stronger effect of symmetry energy as compared to in case
of F$_{3} (u)$ and F$_{4}$.
\par
Since the reaction $^{60}$\textrm{Ca}+$^{60}$\textrm{Ca} is an
extreme case with large isospin asymmetry. To check if the above
predicted effects survive in reactions which are experimentally
accessible, we simulate the reaction of
$^{48}$\textrm{Ca}+$^{48}$\textrm{Ca} for all the different forms
of symmetry energy used in the present study. The reaction
$^{48}$\textrm{Ca}+$^{48}$\textrm{Ca} has been used in many
previous studies \cite{wang09}. The results are shown in fig.
6(a). We find that even for this reaction the transverse flow
shows sensitivity to symmetry energy as well as its density
dependence. In fig. 6(b) and 6(c), we display the time evolution
of maximum density $<\rho_{\textrm{max}}/\rho_{0}>$ and average
density $<\rho_{\textrm{avg}}/\rho_{0}>$, respectively at 100
MeV/nucleon. We find that the density reached is about 2.0 times
the saturation density. Moreover, the maximal density is reached
in the time interval 0-30 fm/c and the effect of symmetry energy
on $<p_{x}^{\textrm{dir}}>$ of particles during this interval
decides the fate of the final value of $<p_{x}^{\textrm{dir}}>$.
Thus transverse flow can address the symmetry energy at densities
about 2.0 times than saturation density.

\par
\section{Summary}
 We have checked the sensitivity of transverse flow towards
 symmetry energy in the Fermi energy as well as at high energies.
 We have found that transverse flow is sensitive to symmetry energy as
 well as its density dependence in the Fermi energy region. We have also shown that the transverse flow can address
 the symmetry energy at densities about twice the saturation density, however it shows the insensitivity towards
 the symmetry energy at densities $\rho/\rho_{0}$ $>$ 2. We have also discussed the
 mechanism for the sensitivity of transverse flow towards symmetry
 energy as well as its density dependence.
 \par
This work has been supported by a grant from Centre of Scientific
and Industrial Research (CSIR), Govt. of India and Indo-French
Centre For The Promotion Of Advanced Research (IFCPAR) under
project no. 4104-1.

\end{document}